\documentclass[a4paper, oneside, twocolumn, notitlepage, 10pt]{extarticle_ecoc}
\usepackage{ecoc}

\newcommand{\mseUnit}{~dB\textsuperscript{2}}

\begin{document}
\selectlanguage{english}    


\title{Machine learning-based EDFA Gain Model Generalizable to Multiple Physical Devices}


\author{
    Francesco Da Ros\textsuperscript{(1)}, 
    Uiara Celine de Moura\textsuperscript{(1)},
    and Metodi P. Yankov\textsuperscript{(1)}
}

\maketitle                  


\begin{strip}
 \begin{author_descr}

   \textsuperscript{(1)} Department of Photonics Engineering, Technical University of Denmark, 2800 Kgs. Lyngby, Denmark,
   \textcolor{blue}{\uline{fdro@fotonik.dtu.dk}} 

 \end{author_descr}
\end{strip}

\setstretch{1.1}


\begin{strip}
  \begin{ecoc_abstract}
   We report a neural-network based erbium-doped fiber amplifier (EDFA) gain model built from experimental measurements. The model shows low gain-prediction error for both the same device used for training (MSE $\leq$ 0.04\mseUnit) and different physical units of the same make (generalization MSE $\leq$ 0.06\mseUnit).
  \end{ecoc_abstract}
\end{strip}


\section{Introduction}
Current communication systems are striving to keep up with the demand for a continuously increasing transmission rate. In parallel to opening new frequency transmission bands and considering spatial division multiplexing, a strong research focus has been directed towards full system optimization. Such a targeted approach has the potential to provide significant enhancement of the system throughput, e.g. by careful optimization of physical layer parameters such as channel power\supercite{IonescuCLEO20}, system margins~\supercite{MahajanJLT20}, and constellation shaping~\supercite{JonesECOC19}. 
Accurate, and more importantly, differentiable models of the transmission subsystems, such as transmission fiber and amplifiers, are critical to perform a realistic end-to-end optimization. 
As accurate model for optical fibers are well-known, a strong focus has been dedicated to amplifier model. Modeling the gain ripples from erbium-doped fiber amplifiers (EDFAs) with gain-flattening filters (GFFs) is discussed in~\supercite{MahajanJLT20}. However, GFFs inherently waste transmission power and removing them has already shown prospects for increased transmission throughput\supercite{IonescuCLEO20}. A number of neural-network (NN) based models for optical amplifiers have been reported recently. NNs have the potential to act as universal approximators, therefore they are promising candidates to provide physical models, and they can be directly trained from experimental data. Additionally, powerful and fully differentiable NN exist, enabling to optimize the full system through gradient descent, i.e. backpropagating the gradient of the error. As such, a number of demonstration of experimentally-trained NN models have been reported for Raman amplifiers\supercite{ZhouJLT06,ZibarJLT19,deMouraOFC2020,ChenOFC20}, hybrid optical amplifiers~\supercite{YeOFC2020}, and EDFAs~\supercite{ZhuOFC20,YouOFC20,IonescuCLEO20,MahajanJLT20}.
In \cite{ZhuOFC20}, good accuracy has been reported but with a rather complex model requiring 90-parallel deep-NN which, in turn, requires a large dataset for training as well as retraining once the amplifier operating condition changes, i.e. different gain levels. Alternatively, the models proposed by \cite{YouOFC20} and \cite{IonescuCLEO20} rely on simpler NNs which generalize to different gain levels without sacrificing accuracy. However, in both cases, the model was trained and tested on a single physical device, which limits its application to optimize practical multi-span systems with different physical devices~\cite{MahajanJLT20}. 

In this work, we propose a simple NN-based EDFA gain model which not only accurately predicts the performance of the specific physical device it is trained on, but it also generalizes well to different physical devices of the same make with negligible loss of accuracy. The EDFA booster model captures gain levels between 10~dB and 22~dB and achieves mean square error (MSE) values below 0.04\mseUnit, and 0.06\mseUnit{ } when testing on the training device or a different device of the same make, respectively. Applications of such a generalizable EDFA gain model to multi-span transmission systems are discussed in\cite{YankovECOC20}. 


\section{Differentiable EDFA model}


\begin{figure}[t]
   \centering
        \includegraphics[width=0.95\linewidth]{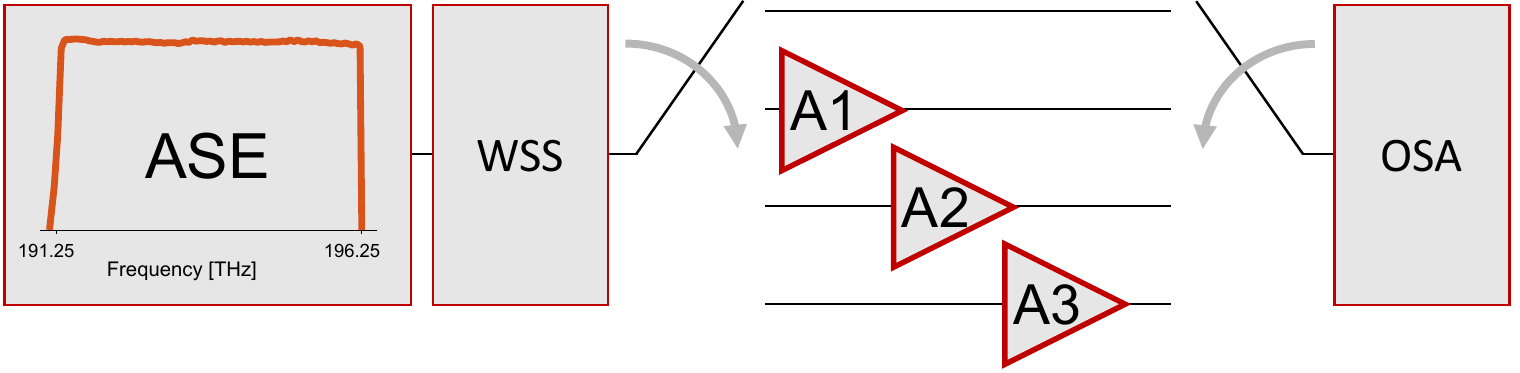}
    \caption{Experimental setup for measuring training and testing datasets}
    \label{fig:figure1}
\end{figure}

\begin{figure*}[t]
   \centering
        \includegraphics[width=0.98\linewidth]{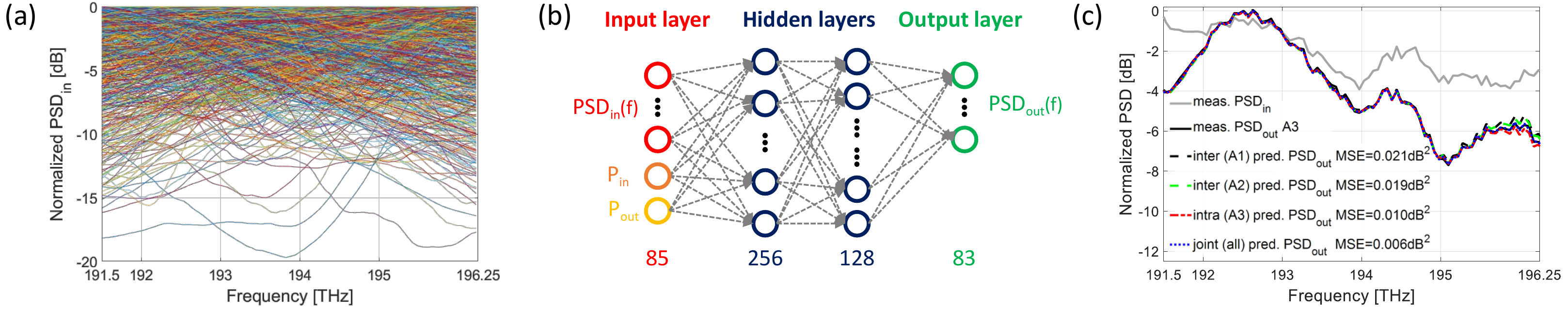}
    \caption{Differentiable EDFA model - (a) Dataset of normalized input power PSDs (PSD\textsubscript{in}), (b) Neural network architecture,  (c) Spectra at the input and the output of A1: input, measured and predicted spectra for models trained under different conditions. }
    \label{fig:figure2}
\end{figure*}

The proposed EDFA model has been built out of an experimentally measured dataset of input and output power profiles and considering three different booster EDFAs of the same make (Keopsys - KPS-STD-BT-C-18-SD-111-FA-FA). The EDFAs do not include GFFs, thus increasing the challenge in achieving an acurate model. The experimental setup employed to gather the datasets used for training and testing the model is shown in Fig.~\ref{fig:figure1}. A fattened amplified spontaneous emission (ASE) source spanning between 191.5~THz and 196.25~THz is spectrally shaped by a wavelength selective switch (WSS). The shaped spectrum is used as input to one of three EDFAs (named A1, A2 and A3) and optical spectra are measured through an optical spectrum analyzer (OSA). A pair of optical switches enables to sequentially record input and output spectra of all three amplifiers. A minimum dataset of 1600 input power spectral density (PSD) profiles for each total input ($P_{in}$) and output power ($P_{out}$) pairs are measured. The input PSDs are shown in Fig.~\ref{fig:figure2}(a) and have been generated considering a channelized power profile (83 equally-spaced frequency channels in the C-band), defined as $p_{n+1} = p_{n} + w_n$, where $p_k$ represents the power of the k-th frequency channel ($k = 1,\dots, 83$), with $p_0$ spanning a 15~dB range, and $w_k \sim \mathcal{N}(0,\sigma^2_W) $ is a Gaussian random variable with zero mean and variance $\sigma^2_W$. The variance of the random walks controls the power excursion of a PSD realization. Excursions up to 20~dB are considered, and the final PSDs have been smoothed using moving averaging filters of different lengths, yielding both sharp and smooth power variations in order to emulate different transmission conditions. This choice is particularly critical in view of considering the model to optimize multi-span systems\supercite{IonescuCLEO20,YankovECOC20}.

The optimized NN architecture is shown in Fig.~\ref{fig:figure2}(b) and consists of an 85-nodes input layer, two nonlinear hidden layers (256 and 128 nodes, respectively) and an 83-nodes output layer. The NN receives in input the normalized 83-channel input PSD (\textit{PSD\textsubscript {in}}), as well as the total input (\textit{P\textsubscript {in}}) and output powers (\textit{P\textsubscript {out}}). The hidden layers consider a RELU activation function and the linear output layer predicts the normalized output PSD (\textit{PSD\textsubscript {out}}). The NN has been implemented in PyTorch, and trained using stochastic gradient descent (Adam algorithm) with MSE between label and predicted spectra as the cost function. The dataset has been split 76\%-24\% between training and testing.

An example of input and output spectra from amplifier 3 (A3) is shown in Fig.~\ref{fig:figure2}(c), together with the predicted output spectra considering four different models: a model trained directly on A3 (intra-EDFA testing), models trained on A1 or A2 (inter-EDFA testing), and finally a model jointly trained on all three amplifiers (joint-EDFA testing). All four models show excellent prediction performance, with intra- and joint-EDFA models performing slightly better than inter-EDFA on the high-frequency region.

\section{Prediction performance}
\begin{figure*}[t]
   \centering
        \includegraphics[width=0.95\linewidth]{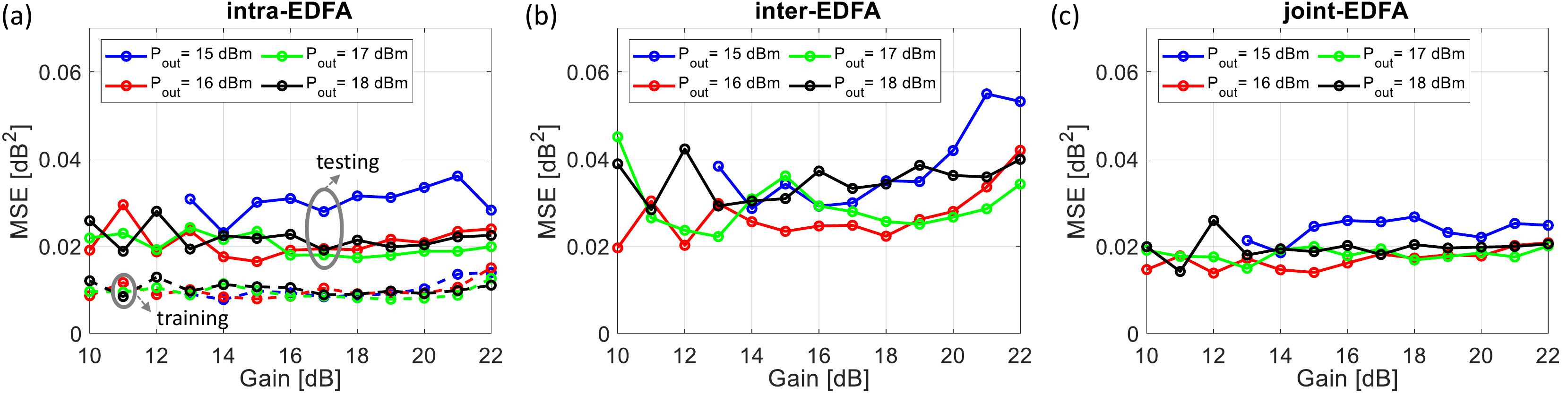}
    \caption{MSE performance - (a) Averaged training and testing MSE for models trained and tested on the same physical unit (intra-EDFA MSE). (b) Averaged testing MSE for models trained on one unit and tested over the other two  (inter-EDFA MSE). (c) Averaged testing MSE for a model jointly trained over all three amplifier (joint-EDFA MSE). }
    \label{fig:figure3}
\end{figure*}

The prediction accuracy of the proposed EDFA models is characterized using the MSE between the predicted output spectrum and the measured spectrum. Fig.~\ref{fig:figure3} shows the MSE averaged over the 83 frequency channels for different gain levels and output powers. Additionally, three test conditions are considered: \emph{intra-EDFA}, i.e. when a model trained with measurements from $Ai,\, i=\{1,2,3\}$ is tested with measurements from $Ai$; \emph{inter-EDFA}, i.e. when a model trained from $Ai$ is tested with measurements from $Aj,\, j\neq i$; and \emph{joint-EDFA}, i.e. when a model jointly trained on all three amplifiers is tested with measurements from each of them. For each scenario, the reported MSE is averaged over the testing dataset for each pairs of (training, testing) EDFAs belonging to the same test condition, i.e. 3  cases for intra-EDFA, 6 cases for inter-EDFA and 3 cases for joint-EDFA. The average number of test profiles is approx. 380 for each pair, leading to test datasets in excess of 1100, 2200, and 1100 input profiles for intra-, inter- and joint-EDFA scenarios, respectively. In Fig.~\ref{fig:figure3}, the MSE values are averaged over the measurement frequency range (limited by our WSS) and the amplifier pair. 
The total input power into the EDFA is varied between -9.3~dBm and 8~dBm with a minimum step size of 0.1~dB, overall targeting gains between 10~dB and 22~dB. In Fig.~\ref{fig:figure3}, the gain are shown gathering values within a $\pm$0.5~dB interval for plotting clarity.

The intra-EDFA performance is shown in Fig.~\ref{fig:figure3}(a), where both testing and training errors are reported. As can be seen, the training error is relatively independent on the gain value and output power with an average MSE of approx. 0.01\mseUnit. Moving to testing, the MSE slightly increases and it shows a higher variance for lower output power and lower gain values. This effect can be related to operating the booster EDFAs at the edge of their operation range, as well as to the slightly lower number of training profiles for $P_{out} = 15$~dBm. Nevertheless, the MSE is well below 0.04\mseUnit, making it comparable to the intra-EDFA test error values reported in~\cite{IonescuICTON19}, and well below \cite{ZhuOFC20}.
In Fig.~\ref{fig:figure3}(b) the generalization performance to different physical units is reported. The inter-EDFA MSE shows an increase compared to the intra-EDFA test, however still below 0.06\mseUnit and with no clear impact from the operation regime (output power and/or gain level). This confirms the effective generalization property of the proposed model both in terms of operation regime of the EDFA and robustness against fabrication tolerances of a specific amplifier unit of a given make.
Finally, in Fig.~\ref{fig:figure3}(c) the MSE performance when a model is jointly trained on measurements for all three amplifiers under test (joint-EDFA) is shown. In this case, the performance are well in-line with the intra-EDFA scenario, with the slightly better performance mainly attributed to the three-times larger training dataset for the joint-EDFA.

\begin{figure}[h]
   \centering
        \includegraphics[width=0.9\linewidth]{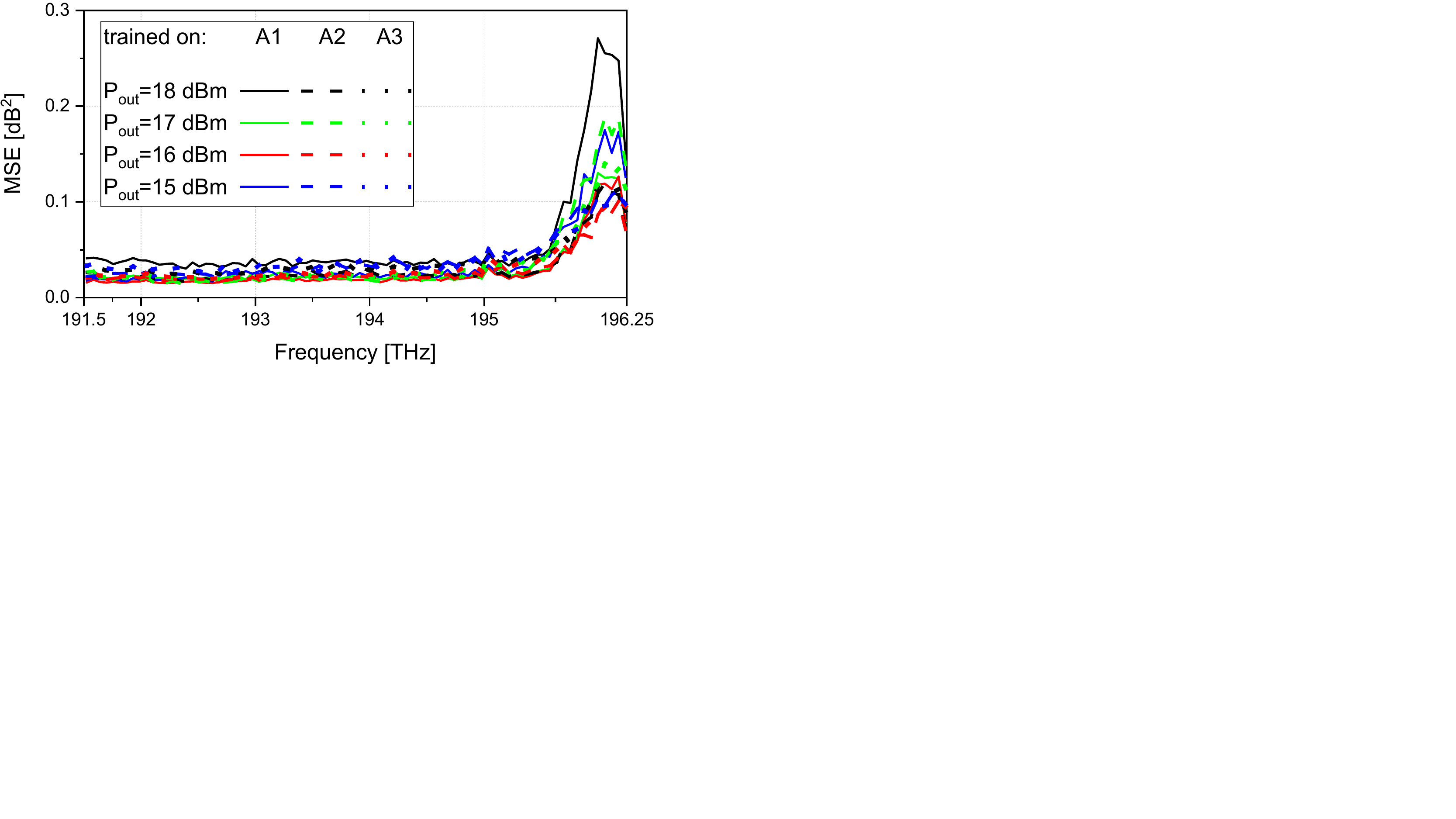}
    \caption{Testing inter-EDFA MSE as a function of the frequency. The MSE is averaged for different gains.}
    \label{fig:figure4}
\end{figure}

The main contribution to the inter-EDFA MSE is investigated in Fig.~\ref{fig:figure4}, where the frequency dependence on the MSE is reported. For this analysis, the MSE is averaged over the 6 inter-EDFA cases and the various gain levels, as Fig.~\ref{fig:figure3}(b) confirmed a negligible gain dependence of the MSE. Regardless of the desired output power, the MSE shows a clear peak in the high frequency range of the measurement bandwidth, i.e. between 195.5~THz and 196.25~THz. 
This clear trend is consistent with the example of Fig.~\ref{fig:figure2}(c), where the predictions for models A1 and A2 deviate more from the measured A3 gain profile, at those frequencies. In general, the inter-EDFA error will be mainly limited by fabrication tolerances between physical units of the same make. This is expected to be related to minor differences between the physical units and to be low enough as  tolerances are constrained by the device specifications. For this work, low enough MSE is achieved such that the model can be representative of multiple devices and used for system optimization~\cite{YankovECOC20}.
%
\section{Conclusions}
A simple, differentiable, experimentally trained, NN-based EDFA gain model is proposed. Accurate gain spectrum predictions generalize to different amplifier operation regimes (gain level and output power), and can be extended to multiple physical devices of the same make. MSE values below 0.06\mseUnit { } are reported for testing over units other than the one used for training the model. The main source of inaccuracy is a slight difference in gain slope in the high-frequency region ($>$195.5~THz). This generalizable EDFA gain model can thus be effectively used as a building block for a full transmission system optimization.

\vspace{-.1cm}
\section{Acknowledgements}
This work is supported by the Villum Foundations (VYI grant OPTIC-AI no.29344), the EU H2020 programme (Marie Sk{\l}odowska-Curie grant no. 754462) and the DNRF CoE SPOC (ref. DNRF123).


\printbibliography
\end{document}